\documentclass[journal,10pt]{IEEEtran}

\usepackage[T1]{fontenc}
\usepackage{amsmath,color}
\usepackage{amsmath}
\usepackage[cmintegrals]{newtxmath}
\usepackage{cite}
\usepackage[utf8]{inputenc}
\usepackage[english]{babel}
\usepackage{verbatim}
\usepackage{syntonly}
\usepackage{textcomp}
\usepackage{amsmath}
\usepackage{upgreek }
\usepackage{makecell}
\usepackage{balance}
\usepackage{varioref}
\PassOptionsToPackage{hyphens}{url}
\usepackage{hyperref}
\usepackage[noabbrev]{cleveref}

\usepackage{tabularx,booktabs}
\newcolumntype{C}{>{\centering\arraybackslash}X} 
\usepackage{lipsum}
\IEEEoverridecommandlockouts
\usepackage{soul}
\usepackage[final]{graphicx}
\usepackage{psfrag}
\usepackage{multicol}
\usepackage{color,comment}
\usepackage{balance}
\usepackage{mathtools}
\usepackage{suffix}
\usepackage{epstopdf}

\renewcommand\[{\begin{equation}}
\renewcommand\]{\end{equation}} 
\definecolor{dkgreen}{rgb}{0,0.3,0}
\definecolor{gray}{rgb}{0.5,0.5,0.5}
\definecolor{violet}{rgb}{0.5,0,0.4}

    \usepackage[compact]{titlesec}
    \titlespacing{\section}{0pt}{2ex}{1ex}
    \titlespacing{\subsection}{0pt}{1ex}{0ex}
    \titlespacing{\subsubsection}{0pt}{0.5ex}{0ex}

\begin{document}
	\title{Aerial Platforms with Reconfigurable Smart Surfaces for 5G and Beyond}
	\author{Safwan Alfattani, Wael Jaafar,  Yassine Hmamouche, Halim Yanikomeroglu, Abbas Yongaçoglu, Ng\d{o}c Dũng Đào, and Peiying Zhu} 

	\maketitle
	
\begin{abstract}
Aerial platforms are expected to deliver enhanced and seamless connectivity in the fifth generation (5G) wireless networks and beyond (B5G). 
This is generally achievable by supporting
advanced onboard communication features embedded in heavy and energy-intensive equipment. 
Alternatively, reconfigurable smart surfaces (RSS), which smartly exploit/recycle signal reflections in the environment, are increasingly being recognized as a new wireless communication paradigm to improve communication links.
\textcolor{black}{In fact}, their reduced cost, low power use, light weight, and flexible deployment  make them an attractive candidate for integration with 5G/B5G technologies. In this article, we discuss 
comprehensive approaches to the integration of RSS in aerial platforms. First, we present a review of RSS technology, its operations and types of communication. Next, we describe how RSS can be used in aerial platforms, and we propose a control architecture workflow. Then, several potential use cases are presented and discussed. Finally, associated research challenges are identified.

\end{abstract}


\sloppy	
	
\section*{Introduction}
As the world steps into the deployment of 5G, researchers continue addressing the challenges of ubiquitous connectivity.
The role of aerial platforms for wireless 
networks is an increasingly important one for meeting the communication requirements of 5G/B5G.
Indeed, the Third Generation Partnership Project
(3GPP) 
has considered aerial platforms to be a new radio access for 5G
(TR 38.811, TR 22.829, and TS 22.125), and several projects have been initiated 
to provide ubiquitous Internet services, such as Google Loon 
 and Thales Stratobus. Due to their 3D-mobility, flexibility, and adaptable altitude, aerial platforms 
can efficiently support the connectivity of terrestrial and aerial users by enhancing their capacity and coverage, or even providing backhaul links. Based on their operation altitudes, we distinguish  two main types of aerial platforms, 
Unmanned Aerial Vehicles (UAVs) and High Altitude Platform Stations (HAPS).
UAVs operate 
at low altitudes of 
few hundred meters and act
as flexible and agile relays or base stations (UxNB) (TR 22.829).
Their use is generally time-limited, ranging from a few minutes to a few hours 
due to limited onboard energy. Alternatively, tethered UAVs
can fly for much longer owing to the continuous supply of power from the tether.
By contrast, HAPS 
operate at higher altitudes of 8 to 50 km above ground 
(TR 38.811), 
with current HAPS projects focusing on the 20 km altitude.
They allow wider coverage areas and longer flight times compared to UAVs (e.g., Google Loon's flight record is 223 days). 
To further reduce costs and make aerial platforms more attractive for 5G/B5G networks, 
researchers are constantly searching for new horizons of
improvements, including fuselage materials, battery, fuel, and solar-cell technologies, as well as enhancing trade-offs in energy efficiency (EE) and communication performance. 

From the communication perspective, passive reconfigurable smart surface (RSS)\textcolor{black}{\footnote{\textcolor{black}{The term \textit{``Smart"} is selected  in accordance with previous 3GPP standards for similar technologies (e.g., TR 22.867, TR 25.842, TR 22.890, TR 36.848). }}} has emerged as a potential substitute for active communication components \cite{Liaskos2018}. RSS is a thin and flexible metasurface equipped with programmable passive circuits to ``reflect'' received signals in a controlled way. An RSS can then positively exploit the propagation environment to enhance 
the quality of a communication link 
between a transmitter and a receiver, known as \textit{RSS-assisted communication}, or to transmit  signals more effectively, identified as \textit{RSS-initiated communication}. 
Efforts have been made 
to evaluate and explore
the benefits of RSS.
For instance, NTT DOCOMO collaborated with Metawave 
and demonstrated an increase in data rate of approximately ten times using RSS at 28 GHz. 
Other notable initiatives include the VisoSurf consortium 
for RSS prototyping,
and Greenerwave, which is already 
marketing RSS that operate at frequencies  up to
100 GHz.

Considering the opportunities presented by RSS technology, 
we discuss in this paper a novel paradigm  where RSS is deployed in 
aerial platforms, and propose a control architecture workflow 
for 
communication, mobility, and sensing.
We then describe potential use cases where RSS-equipped HAPS and UAVs aim to achieve communication objectives, such as coverage extension and low-cost network densification.
Finally, related challenges are presented. To the best of our knowledge, this is the first work that   
presents an in-depth discussion of the use of RSS-equipped aerial platforms, their control architecture, as well as their potential applications and challenges.

\section*{Overview of RSS in Communications}
\subsection*{Background}
Researchers have traditionally focused on
improvements in hardware and transmission techniques
 to enhance  link quality between a transmitter and a receiver.
This research stream 
 has sought ways of mitigating the impact of environmental objects on the quality of wireless signals. 
However, a different approach has emerged whereby the environment between transmitters and receivers can be managed 
\textcolor{black}{to initiate or support wireless communications.} \textcolor{black}{For initiated communications, various ambient energy sources, (RF, solar, etc.), can power ultralow-power electronics to enable sensing functions and backscatter communications through different environmental objects (see \cite{Kimionis2018} and references therein). For assisted communications, the idea of controlling 
surfaces and objects in wireless propagation environments} was 
proposed earlier, around two decades ago. The objective was to mitigate undesired interference and improve the desired signal through the use of frequency selective surfaces (FSS). 
\textcolor{black}{These FSS resemble bandpass or bandstop filters, and they can be integrated into indoor walls. Also, they can be equipped with sensors and diode switches for flexibly controlled response.} 
	

While FSS's key function has been to filter signals and alleviate interference, another vision of utilizing the propagation environment is to combine the scatters of the desired signals and direct them to the targeted users. 
In this context, passive reflectors (a.k.a.,  mirrors) consisting of metallic sheets or high conductivity materials have been designed for indoor and outdoor environments, aiming to enhance the \textcolor{black}{received power. }
For outdoor environments, authors in \cite{Peng2016a} suggested using passive reflectors on top of buildings to focus reflected signals from 
ground base station (gBSs) toward users.
Compared to conventional coverage enhancement approaches,  
such as 
\textcolor{black}{relaying} and gBSs densification,
passive reflectors made of low-cost elements offer flexible deployment, easy maintenance, and improved EE. Nevertheless, passive reflectors are constrained 
by Snell's law of reflection, and once designed, they have a fixed response. Hence, they are inherently unable to accommodate the 
dynamics of wireless environments.
Alternatively, to capture the wireless environment dynamicity, “active” surface technologies were proposed in \cite{Hu2018}, where the walls consist of a massive number of electromagnetically radiating and sensing elements working as large smart antennas, and regarded as an extension to the massive multiple-input multiple-output (MIMO) concept. While passive reflectors utilize material properties to direct electromagnetic waves, active surfaces require integrating electronics, antennas and communication circuits in walls to work as full transceivers. Although they might have better performances, the circuit complexity, material costs, and energy consumption are higher than for passive surfaces.

Inspired by recent advances in the physics of metasurfaces and meta-materials, passive 
RSS technology has 
been presented as an interesting add-on to various applications in wireless communications \cite{Liaskos2018}.
Metasurfaces are distinct man-made 2D meta-materials, consisting of thin layers of metallic/dielectric inclusions. Metasurface units 
 can be arranged in arrays and designed to have peculiar electromagnetic (EM) features with customizable interactions with 
the EM waves.
Indeed, these units can realize abrupt phase shifts and anomalous reflections and refractions, where the phase of the wavefronts can be altered, a process known as  phase discontinuity \cite{Yu2011}. Moreover, the reflected and refracted EM waves can have an arbitrary direction beyond Snell's law, subject to \textit{``the generalized laws of reflection and refraction''} \cite{Yu2011}.
To bypass the fixed response of metasurfaces, digital meta-materials were proposed in \cite{Cui2014}, where the manipulations of the EM waves were reconfigurable by digitally controlling the metasurface units, \textcolor{black}{i.e., switching between different meta-material responses using biased diodes}. 
Subsequently, an advanced layout is proposed in \cite{Liaskos2018},  wherein a programmable metasurface layer was integrated with an Internet-of-Things 
gateway and a control layer  to form a thin film, called ``HyperSurface tile'', coating surfaces and environmental objects.
Hence, the interaction of the surfaces with the EM waves can be programmed and controlled remotely 
via software. 
    \subsection*{Types of Communications with RSS} 
	Given the capabilities demonstrated 
	in physics for controlling and reconfiguring EM waves with meta-materials, wireless communication researchers have started investigating the use of RSS in different ways, particularly for \textit{RSS-assisted communications} and \textit{RSS-initiated communications}.
	
	\subsubsection*{RSS-Assisted Communications}  
	
Aiming for efficient delivery of 
transmitted signals,  
	most research has focused on using RSS as a relay to enhance the received signal 
	at end-users. Different objectives have been considered, such as maximizing  data rates or spectral efficiency, 
	extending the coverage area, mitigating interference, and minimizing consumed energy.
    For instance, the authors of \cite{Wu2019} 
    showed that by  optimizing the RSS phase shifts, 
    quadratic increase  in the signal-to-noise ratio (SNR) is realized, and better spectral efficiency than relays can be achieved.
    To practically substantiate the advantages of RSS-assisted communications, 
    an experimental testbed was implemented in \cite{Tan2018} \textcolor{black}{demonstrating} that RSS can solve the coverage-hole problem caused by 
    obstacles in 
    millimeter-wave (mmWave) networks.
    Additionally, motivated by the EE potential of RSS
    , researchers have  focused on supporting energy-constrained \textcolor{black}{communications}. For instance, authors in \cite{Huang2018} 
    demonstrated that, even with low-resolution phase shifters in RSS-assisted communications, EE can be improved by $40\%$ compared to conventional relaying systems. 

\begin{figure*}[t]
    \centering
\includegraphics[width=0.90\linewidth]{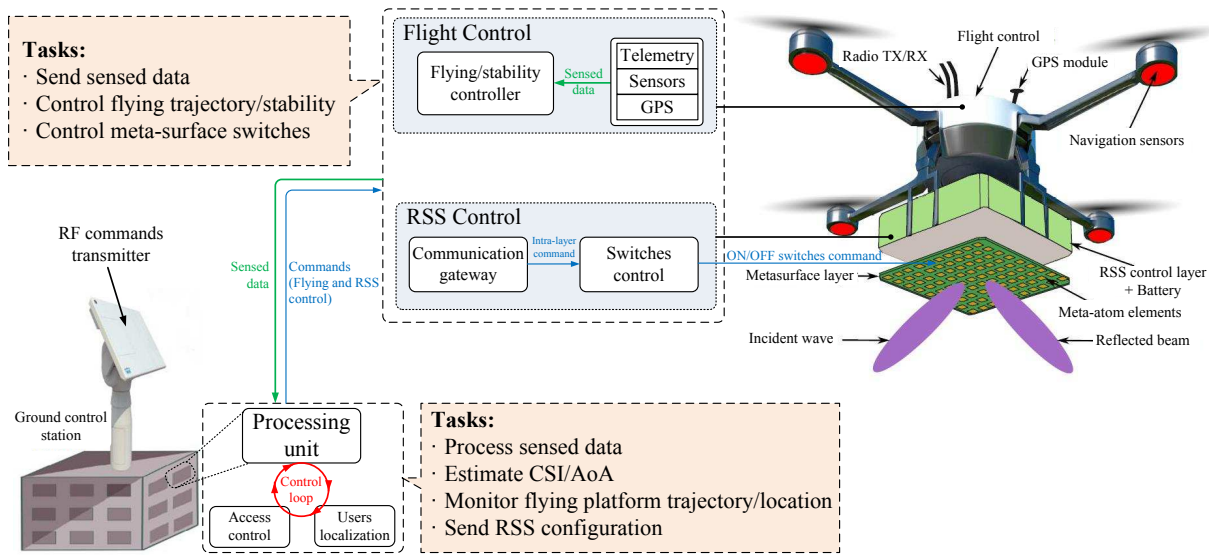}
    \caption{Control architecture of RSS-equipped aerial platforms.}\label{fig:Architecture}
\end{figure*}	
 \subsubsection*{RSS-Initiated Communications}
In contrast to previous works, RSS is being tested as an energy-efficient and low-complex transmitter. The idea is to connect the RSS with a nearby RF source and, through phase shifting, data can be encoded and transmitted. 
While the 
reliability of RSS-initiated communications were analyzed in \cite{Basar2019b}, authors of \cite{Tang} 
validated through experiments the design of a high order modulation RSS-based transmitter, where a digital video file was modulated and transmitted through the RSS 
without requiring filters, mixers, or power amplifiers.


\section*{Aerial Platforms with RSS: Integration and Control Architecture}
The current research works on RSS are focused on its use in terrestrial environments, e.g., RSS equipping buildings facades. However, motivated by the numerous aforementioned RSS capabilities, and the advanced features of aerial platforms, we envision that  the usage of RSS in aerial platforms will offer great flexibility and support for wireless networks. Integrating  RSS on aerial platforms 
can be done in several ways depending
 on the platform’s shape. For instance, the RSS can coat the outer surface of a balloon or an aircraft, or it can be installed as a separate horizontal surface at the bottom of a 
UAV, as shown in Fig. \ref{fig:Architecture}. The reconfiguration of such surfaces can be realized through lumped elements embedded in the reflector units or via the distributed control of some material properties. Examples of lumped elements include PIN diodes, varactor 
diodes, and radio frequency micro-electro-mechanical systems. By tuning the lumped elements, the
resonant frequency of the reflectors can be changed, hence achieving the desired wave manipulation and phase shift. To manage high frequency signals, e.g., mmWave and free-space optical (FSO) communication, tunable materials such as liquid crystal and graphene are used to design the metasurface reflector units.

In order for the RSS-equipped aerial platforms to perform communication functions properly, the channel information needs to be acquired and the RSS should be accurately configured. The control can be managed at two levels: Ground control station and aerial platform, as illustrated in Fig. \ref{fig:Architecture}. 
\subsection*{Ground  Control Station} 
This controller consists mainly of a processing unit that analyzes \textcolor{black}{the} sensed data from the aerial platform, \textcolor{black}{and the} 
access control and users localization conditions, acquired through information exchange with gBSs and/or gateways. 
Given a set of global policies (e.g., flying regulations, power, etc.) and objectives 
(e.g., coverage, \textcolor{black}{SNR}, EE, etc.), the processing unit ensures the joint management of the aerial platform's
flying and communication functions.
Moreover, it estimates the channel state information (CSI) and angle of arrival (AoA) between users and the aerial platform  to determine the best RSS configuration setup.


\subsection*{Onboard Aerial Platform Control}
The onboard aerial platform’s controller consists of two units, namely the \textit{flight control unit}, and the \textit{RSS control unit}. 
The 
former receives motion commands from the ground station, and ensures the platform's stabilization. 
 By contrast, the \textit{RSS control unit} \textcolor{black}{receives the optimized RSS configuration, translates it into a switch control ON/OFF activation map, and applies it to the metasurface layer 
 for directing  incident signals to  targeted directions}.   



\section*{Aerial Platforms with RSS: Potential Use Cases}
To demonstrate the potential benefits of integrating RSS in aerial platforms, we detail here several novel integration use cases.

\subsection*{HAPS with RSS to Support Remote Areas}
	  	 
Connecting sparse users in remote areas via terrestrial networks is deemed to be 
expensive and unprofitable to 
operators due to the high costs of typical 
 gBSs and wired backhauling infrastructure. 
Satellites have been proposed to serve remote areas; however, 
due to high deployment costs, large communication delays, and excessive path-loss, satellites 
are considered as a last resort solution.  
Alternatively, HAPS have been introduced as a cost-effective solution 
for connecting remote users \cite{Alzenad}.
HAPS are quasi-stationary platforms located in the 
stratosphere at a fixed point relative to the earth. 
Since the wind's speed in this region is weak,  
 low power is required to stabilize the platform.
Moreover, the coverage area of a HAPS is greater than that of a gBS. For instance, the International Telecommunication Union (ITU) suggested that the \textcolor{black}{HAPS' coverage radius be} up to 500 km (ITU-R F.1500); however, 
current HAPS projects have less than 100 km coverage radius to achieve high area 
throughputs. 

HAPS are mostly powered by solar/fuel energy, and their flight duration may vary from several hours to a few years.
 The energy consumed by a HAPS is mainly for propulsion, stabilization, and communication operations. Typically, a HAPS is used in communications either as a BS (HAPS-BS) 
 or as a relay station (HAPS-RS). 
Although a HAPS-BS has more capabilities than a HAPS-RS, it involves a higher cost, a heavier payload, and 
an increased energy consumption. 
When solar cells are used, the harvested energy may not be sufficient for operations. 
For this reason, other power sources are added to complement the solar energy. Nevertheless, the 
EE of HAPS is still an open problem, \textcolor{black}{which directly impacts the flight duration}.


\begin{figure}[t]
\psfrag{AAA}[tl][tl][0.8][0]{HAPS-RSS}
\psfrag{BBB}[tl][tl][0.8][0]{Wireless}
\psfrag{NNN}[tl][tl][0.8][0]{Backhaul}
\psfrag{MMM}[tl][tl][0.8][0]{Gateway}
\psfrag{CCC}[tl][tl][0.8][0]{Reflected wireless backhaul}
\psfrag{DDD}[tl][tl][0.8][0]{Flying BS}
\psfrag{GGG}[tl][tl][0.8][0]{Incident/Reflected}
\psfrag{KKK}[tl][tl][0.8][0]{Reflected/Incident}
\psfrag{FFF}[tl][tl][0.8][0]{wireless backhaul}
\psfrag{HHH}[tl][tl][0.8][0]{Rural isolated users}
\psfrag{EEE}[tl][tl][0.8][0]{Wired backhaul}
\psfrag{III}[tl][tl][0.8][0]{Core network}
\includegraphics[width=0.9\columnwidth]{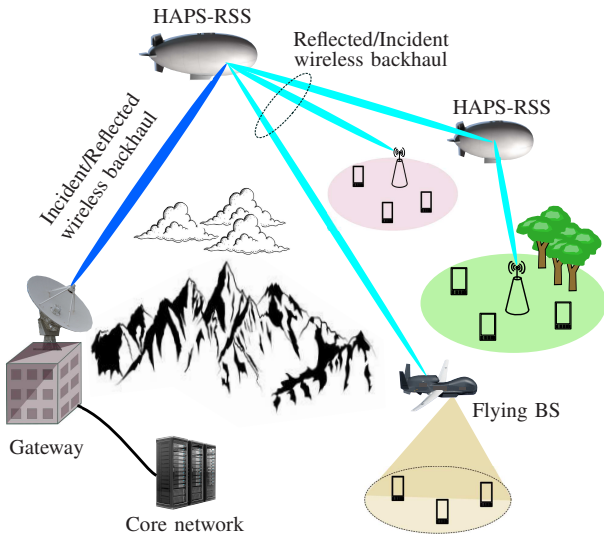}
\caption{Backhauling support of isolated users in remote rural areas using HAPS-RSS.}
\label{Fig:HAP}
\end{figure}
In designing a HAPS deployment, low energy consumption, light payload, and reliable communication need to be achieved.
To this end, motivated by the benefits of RSS, we propose using a HAPS equipped with RSS
(HAPS-RSS) for wireless traffic backhauling from remote area 
BSs.
The envisioned scenario is depicted in Fig. \ref{Fig:HAP}, where 
BSs handle the traffic of a few clustered users and transmit the traffic to the HAPS. Through the HAPS' RSS, the received signals are smartly ``reflected'' 
toward a gateway station, which is connected to the core network. 
If a gateway station is not within the HAPS coverage range, the RSS can be configured to ``reflect'' the signals toward another HAPS and thereby reach the gateway station in a dual or multi-hop fashion.

Using HAPS-RSS has several advantages. 
First, it reduces the used power for communication functions, since directing signals can be realized in a nearly-passive way, without RF source or power amplifiers, and only minimum amount of power is required for the RSS control unit. 
Second, since RSS is made of thin and lightweight materials, the HAPS payload is lighter even if a large number of reflectors is deployed. Third, given the reduced communication components and payload, the required stabilization  
energy 
is minimized. 
The simplification in communication circuits and minimization of consumed energy yield to extended flight duration and reduced HAPS deployment costs.

 The performance of RSS-assisted communications strongly depends on the number of reflectors. 
 The dimensions of a metasurface reflector unit (a.k.a.,  meta-atom)  are proportional to the wavelength \cite{Liaskos2018}. 
 Typically, a HAPS is a giant aircraft with a length between 30 and 200 m,
 which has the potential to accommodate a large number of reflectors, especially in high frequencies. 
 The large number of reflectors would have a significant impact on the directivity gain, hence 
 better SNR performance than HAPS-RS can be achieved \cite{Wu2019}. 
In terms of energy consumption, the switches on the metasurface layer consume very low power. A recent experiment showed that each meta-atom consumes 0.33 mW when its switch diode is set to ON \cite{Tang2019}. Thus, for a HAPS with 10,000 meta-atoms, 
the maximal power consumed by the metasurface layer when all switches are activated would be about 3.3 W,
which constitutes a very promising result. 
Hence, HAPS-RSS is expected to achieve a better EE than HAPS-RS \cite{Huang2018}.  
Finally, it is worth noting that RSS supports full-duplex communication \cite{Wu2019}. Indeed, unlike relays experiencing high noise and residual loop-back self-interference, which are issued from active RF components, RSS bypasses these constraints by only exploiting the intrinsic properties of the reflecting material. Thus, HAPS-RSS would be more spectrally efficient than HAPS-RS.
Table \ref{tab1} summarizes the main differences between HAPS-RS and HAPS-RSS communications. 

\begin{table}
		\caption{Comparison between HAPS-RS and HAPS-RSS for communication}
	\label{tab1}
	\footnotesize
	\begin{tabular}{|p{50pt}|p{81pt}|p{81pt}|}
    \hline
		\makecell{\textbf{Aspect}}	 & \makecell{\textbf{HAPS-RS}} & \makecell{\textbf{HAPS-RSS}} \\ 
		\hline
		\makecell{\textbf{Energy}\\ \textbf{Consumption}} & \makecell{\textbf{Medium} \\(Process, amplify,\\ and forward signals)}
		& \makecell{\textbf{Very low} \\(RSS control)}
		\\ 
		\hline
		\makecell{\textbf{Payload}} & \makecell{\textbf{Heavy} (Complex \\circuits: converters,\\ mixers, amplifiers, \\and antennas)}
		& \makecell{\textbf{Light} (RSS designed as \\a thin sheet film coating \\HAPS outer surface)}
		\\ 
		\hline
		\makecell{\textbf{No. of}\\ \textbf{Commun.}\\\textbf{Units}} & \makecell{Limited by frequency \\and energy constraints, \\and by allocated space \\for circuits.} & \makecell{Limited by HAPS size \\and frequency band.} \\ 
		\hline
		\makecell{\textbf{Performance}\\ \textbf{(directivity}\\\textbf{gain)}} & \makecell{Good performance,\\ low EE} & 
		\makecell{High EE, performance \\depends on RSS size\\ and reconfiguration\\ capability} \\ 
		\hline
       \makecell{\textbf{Spectral}\\ \textbf{efficiency}} & \makecell{Good in full-duplex, \\but limited by noise \\and self-interference.} & \makecell{High, always \\full-duplex and no \\noise or interference.}\\
       \hline
		\makecell{\textbf{Cost}} & \makecell{\textbf{High} (Several active\\ components, short flight \\duration)}
		& \makecell{\textbf{Low} (Low-cost RSS,\\ extended flight duration)}
		\\ 	  	
	\hline
	\end{tabular}
\end{table}

\subsection*{UAVs with RSS to Support Terrestrial Networks}

\begin{figure}
\vskip -0.4cm
\hskip -0.7cm 
\begin{center}
\psfrag{FCDSZ}[tl][tl][0.65][0]{A swarm of UxNB-RSS}
\psfrag{OOO}[tl][tl][0.65][0]{Isolated user}
\psfrag{FFF}[tl][tl][0.65][0]{Impenetrable blockages}
\psfrag{PPP}[tl][tl][0.65][0]{gBS with directional antennas}
\psfrag{HTC}[tl][tl][0.65][0]{toward UxNB-RSS}
\psfrag{JJJ}[tl][tl][0.65][0]{gBS with advanced 3D beamforming}
\psfrag{VVV}[tl][tl][0.65][0]{Ground control}
\psfrag{III}[tl][tl][0.65][0]{station}
\psfrag{GSED}[tl][tl][0.8][0]{}
\includegraphics[scale=0.14]{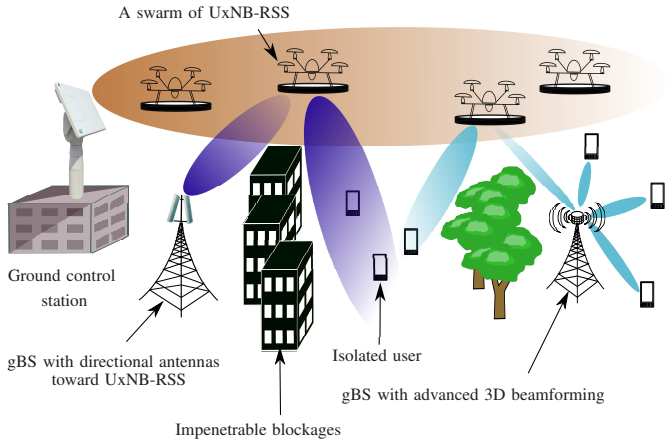}
\caption{Communication support of terrestrial networks with blockages using UxNB-RSS.}\label{Fig:RSS-UAV}
\end{center}
\vskip -0.4cm
\end{figure}

RSS can be combined with UAVs and their attendant benefits, namely, agility, flexibility, and rapid deployment,  to assist terrestrial cellular networks 
in a cost-effective manner.
Typically, RSS can be carefully mounted on a swarm of UAVs (UxNB-RSS) to create an intermediate reflection layer between gBSs and isolated users, as illustrated in Fig. \ref{Fig:RSS-UAV}. In this way, the UxNB-RSS allows a smooth mechanical movement of the RSS layer, while the RSS enables digitally-tuned reflections of incident signals to improve service for isolated users. Hence, besides its low manufacturing cost, UxNB-RSS can be rapidly deployed to increase operators' revenue per user by filling coverage holes and meeting users' high-speed broadband needs.

Coverage holes or blind spots can occur for various reasons, such as disruption of some gBSs due to maintenance, failures, or natural disasters. Also, 
future communication networks are exploring the use of higher frequencies with wider spectrum bandwidths, such as mmWave and FSO, to tackle the increasing throughput
demands and the spectrum scarcity problem. A major challenge hindering their widespread use is their vulnerability to blockages in the propagation environment
, which requires perfect line-of-sight (LoS) links but may create coverage holes in regions with impenetrable blockages. In such a context, 
\textcolor{black}{UxNB-RSS, which has a panoramic view of the environment, can provide full-angle $360^\circ$ reflection and thus can reduce the number of reflections compared to terrestrial RSS \cite{lu2020enabling}.} This approach can also be adopted to deliver new signals to users served by a heavily loaded gBS, and thus help in balancing the traffic of congested cells.

In the absence of channel knowledge, UxNB-RSS can be used for spatial diversity to combat channel impairments (e.g., fading and path-loss). A typical user can actually receive several copies of the desired signal over different transmission paths generated by reflections through the swarm of UxNB-RSS, which enhances the reliability of the reception. Alternatively, fading can be seen as a source to increase the degrees of freedom. A high-rate data stream can be divided into many lower-rate streams and transmitted over different transmission paths using the swarm of UxNB-RSS. Thus, 
the user can retrieve independent streams of information with sufficiently different spatial signatures, resulting in improved data throughput. 
It is also worth noting that RSS can absorb a certain amount of the impinging signals and leverage them to recharge the UxNB-RSS. 

\subsection*{Tethered Balloons with RSS to Support Terrestrial/Aerial Users}	
\begin{figure}
\vskip -0,1cm 
\begin{center}
\psfrag{BBB}[tl][tl][0.7][0]{Tether}
\psfrag{CCC}[tl][tl][0.7][0]{RSS controller}
\psfrag{GGG}[tl][tl][0.7][0]{Gateway}
\psfrag{DDD}[tl][tl][0.7][0]{Core network}
\psfrag{VVV}[tl][tl][0.7][0]{UAV-UE}
\psfrag{FFF}[tl][tl][0.7][0]{TBAL-RSS}
\includegraphics[scale=0.150]{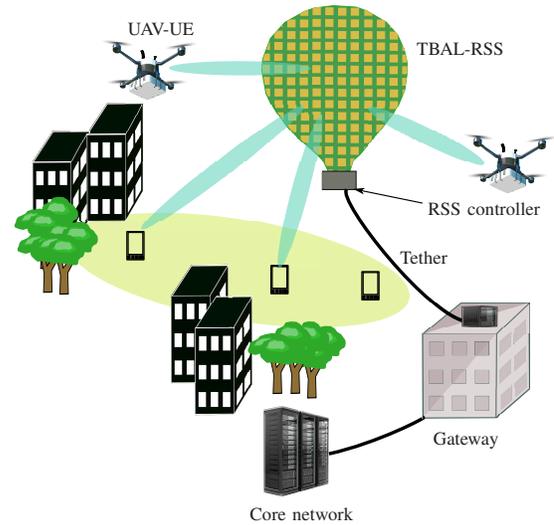}
\caption{Communications support of terrestrial and aerial users using TBAL-RSS.
}\label{Fig:balloon}
\end{center}
\vskip -0.6cm
\end{figure}

UxNB has been viewed as a candidate solution for several wireless challenges, such as coverage and capacity extension. 
Although they have promising advantages, such as flexible placement and LoS communication links, there are also several limitations that militate against using UxNB.
First, the limited onboard energy 
makes UxNB exclusively suitable for short-term deployments. Second, backhauling from a UxNB is typically through a wireless link, which has limited reliability and capacity  compared to wired backhauling links. 
To bypass these limitations
, a tethered UxNB 
\textcolor{black}{has been recently introduced.} 
The tether provides the UxNB with both data and power. Thus, longer flight times and reliable backhaul links are achieved. However, a tethered UxNB consumes a considerable amount of energy for flying, hovering and transmitting signals
, which yield to increased deployment and operational costs.
A more cost-effective alternative \textcolor{black}{that} we propose here is using a tethered balloon equipped with RSS (TBAL-RSS), 
as depicted in Fig.~ \ref{Fig:balloon}.

Inspired by recent works and prototypes that use RSS as  transmitters \cite{Basar2019b,Tang}, 
we point out that TBAL-RSS can be used 
as a low-cost wireless access point to connect or enhance the capacity of urban users. 
By illuminating the outer surface of the TBAL-RSS with an RF feeder
, the information provided by the tether data link can be encoded and transmitted. Thus, the RSS controller adjusts the phases of the reflected signals to generate different beam directions that can serve multiple users.
Compared to a tethered UxNB, TBAL-RSS has several advantages. Unlike 
a small tethered UxNB, a TBAL 
is large and can accommodate a higher number of reflectors. 
Also, compared to typical gBSs, network densification with TBAL-RSS in urban environments is more feasible and cost-effective. 
Since TBAL-RSS provides near-field communications with strong LoS links to users, it is expected to provide better spectral efficiency than MIMO gBSs,
considered as a far-field communication system with \textcolor{black}{many NLoS links, 
high interference, and pilot contamination.} 
Furthermore, unlike down-tilted antennas in gBSs, the TBAL-RSS can sustain
reliable connectivity to \textcolor{black}{UAV users (UAV-UEs)} such as delivery drones. 
Thus, both terrestrial and aerial users can be served simultaneously.

Moreover, a TBAL does not continuously drain power for lifting and hovering, as it 
uses lifting gases.
Although the altitude of a TBAL is limited by the tether's length, some flexibility in its position is available by controlling the tether's direction to offer better coverage in certain spots.
Besides, a backhaul fiber optic link along the tether can be installed to guarantee the TBAL's connectivity to the core network. A processing unit is also placed at the TBAL's launching point to control its movement, determine the targeted users, and send the optimum RSS configuration. 

{\color{black}
\section*{Case Study}
\begin{figure}
	\centering
	\includegraphics[width=1.05 \linewidth]{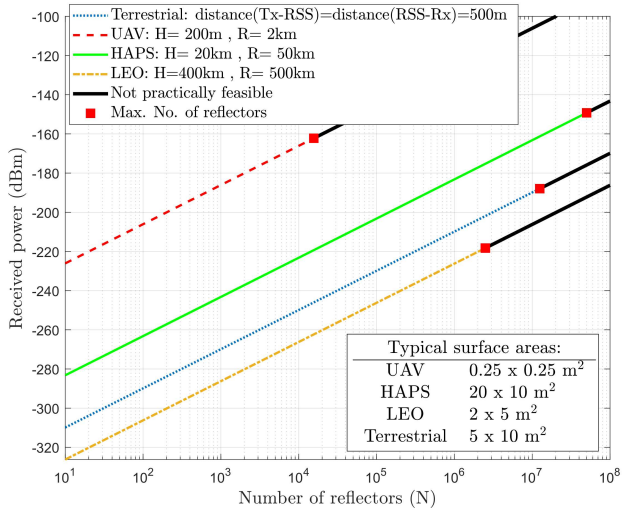}
	\caption{Received power vs. number of reflectors (different platforms).}
	\label{Fig:Pr_vs_N}
\end{figure}
To emphasize the potential gains when using RSS-equipped aerial platforms, we present a case study where the communication between 
a terrestrial transmitter and a terrestrial receiver
is assisted by an RSS-equipped aerial platform, namely, a UAV, a HAPS, or a low earth orbit (LEO) satellite.
Each platform has a typical coverage footprint and deployment altitude, where typically higher altitude platforms have wider footprints, i.e., larger coverage radius ($R$). For instance, a UAV may have $R=2$ km, while a HAPS has $R \in [50,100]$ km.
For the sake of simplicity, we assume that the aerial platform is located at the middle distance between the transmitter and receiver
, which are located at opposite edges of the coverage footprint. Moreover, platforms have different sizes that delimit the practical RSS mounting areas on them. Given the RSS area and operating frequency, the maximum number of reflectors on the RSS can be bounded.}

\textcolor{black}{Following the scattering paradigm-based link budget analysis conducted in \cite{Alfattani2020}, where perfect RSS phase shifting and no reflection losses are assumed,
we compare in Fig. \ref{Fig:Pr_vs_N} the received powers when the communication is assisted by 
RSS in aerial platforms or in terrestrial environments.
The characteristics of each platform are shown in the legend boxes.
We set  the transmit power and the frequency to 40 dBm and 30 GHz, respectively. Also, we assumed unit antenna gains at both transmitter and receiver, and a path-loss exponent equal to 4 for the terrestrial 
environment.
We notice that at a fixed number of reflectors, the UAV 
achieves the best performance. This is expected due to 
strong LoS  links compared to the terrestrial RSS (where blockages degrade the communication links), and the other aerial platforms (where longer distances are traveled by the transmitted signal). 
However, since the UAVs' RSS area are limited and some platforms can host a larger number of RSS reflectors (red squares), HAPS become better than UAVs. 
Ultimately, the HAPS is preferred due to its achieved performance and sustainability.}


\section*{Related Research Challenges}
As the research of RSS in wireless communications is still in its infancy, several challenges and open problems remain to be addressed. In the following, we discuss some key issues.

\subsection*{Accurate RSS and Aerial Channel Models}
Most of the existing research works consider RSS with perfect manipulation of the EM wave and ideal phase shifts. However, there is a paramount need for practical RSS models that consider the configuration capabilities and the reliability 
with different communication frequencies and diverse numbers and sizes of RSS units. 

Furthermore, although several air-to-ground channel models have been developed, integrating RSS in aerial platforms is a new direction that requires the investigation of new aerial models that consider  
RSS capabilities with  aerial platforms properties.
Some important factors that 
could affect path-loss models need to be analyzed, such as reflection loss, correlation between RSS units, 
and atmospheric attenuation. \textcolor{black}{  Moreover, the mobility of aerial platforms may cause fluctuations and beam misalignments   
which have to be considered, especially for HAPS and high frequency communications.}


\subsection*{RSS Deployment on Aerial Platforms' Surfaces}
Aerial platforms have different surface shapes and types. They might be 
flat, like the bottom of a UAV, or curved, like the outer surface of an aircraft or balloon.
How such an RSS deployment may affect metasurface configuration and performance has not been investigated yet.

\subsection*{Channel Estimation and RSS Control Update}
One of the fundamental assumptions in most current RSS research is the availability of CSI, which is essential for performing perfect reflection. However, gaining such knowledge within the channel's coherence time  in typical dynamic wireless environments with mobile platforms is still an open problem. 
One solution to this is by incorporating low-cost sensors in metasurfaces \textcolor{black}{\cite{Kimionis2018}}, which frequently sense the environment and send measurements to the controller. The latter computes and sends back the updated optimal RSS configuration. 
Alternatively, \textcolor{black}{the reciprocity of the channel can be exploited to estimate the CSI for pre-determined phase shifting configurations, or} machine learning techniques, such as supervised learning \textcolor{black}{and} reinforcement learning, \textcolor{black}{can be leveraged} to find the optimal RSS configuration.  
However, such techniques require large amounts of data and/or long training times. 
Hence, 
the delay impact 
due to distance and CSI estimation process merits further investigation.



\section*{Conclusion}
In this article, we discussed the potential integration of the recent RSS technology into aerial platforms as a novel 5G/B5G paradigm. First, we provided an overview of RSS technology, its operations, and types of communication. Then, we proposed a control architecture workflow for the integration of RSS in aerial platforms. Potential use cases were discussed where different types of aerial platforms, namely HAPS, UAVs and TBALs, were exploited. 
Integrating RSS in aerial platforms provides several advantages, including energy efficiency, lighter payload, and lower system complexity, which results in extended flight durations and more cost-effective deployment to support wireless networks. However, several 
issues need to be tackled for the smooth integration of this novel paradigm. 




\bibliographystyle{ieeetr}

\balance

\section*{Biographies}
\small{
\noindent \textbf{Safwan Alfattani} [S] (smalfattani@kau.edu.sa) received his B.S. degree 
from King AbdulAziz University, Saudi Arabia. He received his M.Sc. degree 
from University of Ottawa, where he is currently working toward his Ph.D. His research interest include aerial networks, reconfigurable smart surfaces, and IoT networks.

\noindent \textbf{Wael Jaafar} [SM] (waeljaafar@sce.carleton.ca) is an NSERC Postdoctoral Fellow at Carleton University. His research interests include wireless communications, edge caching/computing, and machine learning. He is the recipient of prestigious grants such as NSERC Alexandre Graham-Bell (BESC) and FRQNT internship scholarship.


\noindent \textbf{Yassine Hmamouche} [S] (hmamouche@ieee.org) received the Ph.D. degree in telecommunications from IMT Atlantique, Brest, France, in 2020. Prior to joining academia, he had extensively worked for eight years in the Telecom industry as a Radio Network Planning Engineer, and then as a Project Manager for several projects spanning over different technologies.

\noindent \textbf{Halim Yanikomeroglu} [F] (halim@sce.carleton.ca)  is a professor 
at Carleton University, Canada. His research interests cover many aspects of 5G/5G+ wireless networks. His collaborative research with industry has resulted in 37 granted patents. He is a Fellow of IEEE, the Engineering Institute of Canada (EIC), and the Canadian Academy of Engineering (CAE), and he is a Distinguished Speaker for IEEE Communications Society and IEEE VT Society.

\noindent \textbf{Abbas Yongaçoglu} [LM] (yongac@uottawa.ca) is an Emeritus Professor at the University of Ottawa.
His area of research is wireless communications and signal processing. He is a Life Member of IEEE.

\noindent \textbf{Ng\d{o}c Dũng Đào}  (ngoc.dao@huawei.com) is a principle engineer
of Huawei Technologies Canada. His research interest covers 5G/5G+
mobile networks. He is a series editor of IEEE Communications
Magazine.

\noindent \textbf{Peiying Zhu} [F] (peiying.zhu@huawei.com) is a Huawei Fellow. She is currently leading 5G wireless system research in Huawei
Technologies Canada. The focus of her research is advanced wireless access technologies with more than 150 granted patents.
}

\end{document}